\documentclass[12pt]{article}
\usepackage{graphicx}

\def \to {\rightarrow}
\def \beq {\begin{equation}}
\def \eeq {\end{equation}}
\def\bfsig{\mbox{\boldmath$\sigma$}}
\def \ba {\begin{eqnarray}}
\def \ea {\end{eqnarray}}

\def \jpsi {J/\psi}
\def \twopi {\pi^+ \pi^-}

\def \smatr {\langle f | S | i \rangle}

\begin{document}
\begin{flushright}
AS-ITP-2001-016 \\
hep-ph/0109055
\end{flushright}
\vskip 10mm
\begin{center}
{\Large Exclusive Decay of $1^{--}$ Quarkonia and $B_c$ Meson into a Lepton Pair
Combined with Two Pions}\\
\vskip 10mm
J. P. Ma   \\
{\small {\it Institute of Theoretical Physics, Chinese Academy of Sciences,
Beijing 100080, China }} \\
~~~ \\
Jia-Sheng Xu \\
{\small {\it China Center of Advance Science and Technology
(World Laboratory), Beijing 100080, China }} \\
{\small {\it and Institute of Theoretical Physics, Chinese Academy of Sciences,
Beijing 100080, China }}
\end{center}

\vskip 1 cm


\begin{abstract}
We study the exclusive  decay of $J/\Psi$, $\Upsilon$ and $B_c$
into a lepton pair combined with two pions in the two kinematic
regions. One is specified by the two pions having large momenta,
but a small invariant mass. The other is specified by the
two pions having small momenta. In both cases we find that in the
heavy quark limit the decay amplitude takes a factorized form, in
which the nonperturbative effect related to heavy meson is
represented by a NRQCD matrix element. The nonperturbative effects
related to the two pions are represented by some universal
functions characterizing the conversion of gluons into the pions.
Using models for these universal functions and chiral perturbative
theory we are able to obtain numerical predictions for the decay
widths. Our numerical results show that the decay of $\jpsi$ is at
order of $10^{-5}$ with reasonable cuts and can be observed at BES
II and the proposed BES III and CLEO-C. For other decays the
branching ratio may be too small to be measured.
\end{abstract}

\vfill\eject\pagestyle{plain}\setcounter{page}{1}




\section{Introduction}
$50 \times 10^6$ $\jpsi$ events have been collected with the upgraded
Bejing Spectrometer (BES II) at Beijing Electron Positron Collider
(BEPC), and several billions $\jpsi$ events will be collected with the
proposed BES III at BEPC II and CLEO-C at modified Cornell Electron
Storage Ring (CESR)\cite{zhao,cleoc}.
Furthermore, about $4 ~{\rm fb}^{-1}$ $b{\bar b}$ resonance data are
planned to be taken at CLEO III in the year prior to conversion to low
energy operation (CLEO-C)\cite{cleoc}.
With these data samples various decay modes of
$\jpsi$ and $b {\bar b}$ resonances can be studied with high statistics.
In this paper we propose to study the exclusive decay of $1^{--}$ quarkonia
and $B_c$ into a lepton pair and
a pion pair. We consider two limited cases in the kinematic
region. One is specified by the pion pair having a large total momentum
and a small invariant mass. In this case, the pions are hard.
The other is specified that the pion pair having a small momentum, i.e.,
the pions are soft. In these decays the two pion system is produced
through conversion of gluons into the two pions. Because of isospin
symmetry conversion of gluons into one pion is highly suppressed.
In a two-pion system the two pions can be in an isospin singlet, the
conversion is allowed. Hence these decays will provide valuable
information how unobservable gluons, as dynamical degrees of freedoms
of QCD, are converted into observable hadrons.

\par
In the case that two-pion system has an invariant mass $m_{\pi\pi}$ which is much
smaller than the mass of heavy
meson and has a large total momentum,
the decay amplitude takes a factorized form in the heavy quark limit,
in which the nonperturbative
effect related to heavy meson is represented by a
non-relativistic QCD (NRQCD) matrix element\cite{nrqcd}, and that
related to the two pions is represented by a distribution amplitude
of two gluons in the isoscalar pion pair which is defined with
twist-2 operators. The two gluons are hard in the kinematical
region, their emission rate can be calculated
with perturbative QCD. The same distribution amplitude also
appears in the predictions for
productions of two pions in
exclusive processes $\gamma+\gamma^*\to \pi+\pi$\cite{diehl,D2,KMP},
$\gamma^* +h \to h +\pi+\pi$\cite{PL1,L1}, and the radiative decay
of $1^{--}$ heavy quarkonium\cite{ma-xu}, where the amplitudes
can be factorized in certain kinematic region. Besides these processes,
the decays studied here will provide another way
to study the nonperturbative mechanism
of how gluons, which are fundamental dynamical freedoms of
QCD, are transmitted into the two pions.
Furthermore, for $\gamma+\gamma^*\to \pi+\pi$, at the tree-level,
only the distribution amplitude of quark appears
in the scattering amplitude,  the distribution amplitude of
gluon appears at loop levels or through evolution of distribution
amplitudes\cite{diehl,D2,KMP}, while for $\gamma^* +h \to h +\pi+\pi$,
at the tree-level, the distribution amplitude of quark as well as that
of gluon contribute to the scattering amplitude, but the produced
charged pion pair is dominantly in an isospin $I=1$ state\cite{PL1,L1}.
This may make it difficult to extract the distribution amplitude of
gluon from corresponding experimental data, because the two pions
produced through gluon conversion are in a $I=0$ state.
For the decays  studied in this paper
and the radiative decay of $1^{--}$ heavy quarknonium
to two pions\cite{ma-xu}, only the distribution amplitude of gluon
appears at the tree-level and the produced two pions are dominantly in a
$I=0$ state. This makes the extraction of gluon content of a
two-pion system relatively easier in experiment.
Of course, comparing with the radiative decay of $1^{--}$ heavy
quarkonium into two pions in the same kinematic
region, the leptonic decay of $1^{--}$ heavy quarkonium to
two pions is suppressed by the fine structure constant $\alpha$,
but the final state in the latter case is more clearer and
can be detected with higher efficiency.
With the model for the distribution amplitude of gluon, developed in
\cite{KMP,L1}, we obtain numerical predictions for the
branch ratio of the decay in the considered kinematic
region. Our results show that
the decay mode of $\jpsi$ in the considered
kinematic region is  certainly
observable at BES II and the proposed BES III and CLEO-C. For other decays
the branching ratios may be too small to be measured.

\par
In the case with tow soft pions, it has been shown that
the decay amplitude of $J/\Psi$ and of $\Upsilon$  also takes a
factorized form in the heavy quark limit\cite{jpma}.  In the decay
amplitude, the nonperturbative effect related to heavy quarkonium
and that related to pion pair can be separated, the former is still
represented by a NRQCD matrix element, while the later is represented
by a matrix element of a correlator of electric chromofields which
characterizes soft gluons transition into the pion pair. This result is
nonperturbative. For $B_c$ decay one can generalize the
approach and obtain a factorized form for the decay amplitude,
where the same correlator appears.
Since the matrix element of the correlator of
electric chromofields between the vacuum state and the two-pion state is
unknown, no numerical prediction of the decay is given in \cite{jpma}.
In this paper, we develop a model for the matrix element
of the correlator of electric chromofields and give numerical
predictions for the leptonic decays
$\jpsi, \Upsilon (1S)$ and $B_c$ into a soft pion pair.
Numerical results are obtained in the considered region
of kinematics and show that
the decay mode of $\jpsi$
is observable at BES II and at the proposed BES III and CLEO-C.

\par
The paper is organized as the following:
In Sec. 2  we study the decays of $\jpsi$ and
$\Upsilon (1S)$ into two hard pions combined with a lepton pair,
where we give a detailed derivation of the factorized
amplitude of the decay. Numerical results for
the decays are presented.
In Sec. 3  the decays of $\jpsi$ and
$\Upsilon (1S)$ into two soft pions combined with a lepton pair are studied,
a model for the matrix element
of the correlator of electric chromofields is developed,
and numerical results for the decay are given.
Sec. 4 is devoted to the study of the decays of
$B_c$. We summarize our work in Sec. 5.

\par
In this paper, we take nonrelativistic normalization for
the heavy meson states and for heavy quarks, and we take
the pion pair to be of a $\pi^+$ and a $\pi^-$. Using
isospin symmetry one can easily obtain results for
a pair of $\pi^0$'s.


\section{Leptonic decays of $\jpsi$ and $\Upsilon (1S)$ to
two hard pions}
We study the exclusive decay in the rest frame of
$\jpsi$:
\beq
 \jpsi (P) \to l^+ (p_1) + l^- (p_2) + \pi^+ (k_{\pi^+})
 + \pi^-(k_{\pi^-}),
\eeq
where $l= e, \mu$, the momenta are indicated in the brackets.
We denote $k = k_{\pi^+} + k_{\pi^-}, q = p_1 + p_2$ and
$m^2_{\pi\pi}=k^2$. We consider the kinematic region
where $\vert{\bf k}\vert\gg m_{\pi\pi}$ and $m_{\pi\pi}\ll M_{\psi}$.
At leading order of QED, the S-matrix element for the decay is
\beq
\smatr =  - i Q_c e^2 L_{\mu} \cdot \frac{1}{q^2}
\int d^4 z e^{i q \cdot z} \langle
\twopi | \bar{c} (z) \gamma^{\mu} c(z) | \jpsi
\rangle ,
\eeq
where $Q_c$ is the electric charge of c-quark in unit of $e$,
$c(z)$ is the Dirac field for c-quark, and
\beq
L_{\mu} = \bar{u} (p_2) \gamma_{\mu} v (p_1).
\eeq
At leading order of QCD,
two gluons are emitted by the c- or $\bar{c}$-quark, and these two
gluons will be transmitted into the two pions. Using Wick theorem
we obtain:
\ba
\smatr &=&  \frac{i}{2} ~\frac{1}{2} \delta^{a b}
Q_c e^2 g_s^2 L_{\mu} \cdot \frac{1}{q^2} \nonumber \\
&& \times \int d^4 z ~d^4 y
~d^4 x ~d^4 x_1 ~d^4 y_1
~e^{i (q \cdot z + k_2 \cdot y) }  \nonumber \\
&& \times ~\langle 0 | \bar{c}_j (x_1) c_i (y_1) | \jpsi \rangle \
 ~\langle \twopi |  G_{\mu_1}^{a} (x) G_{\nu_1}^{b} (0) |0
\rangle
\nonumber \\
&& \times ~[ \delta^{4} (x - x_1) \delta^{4} (z - y_1)
\gamma^{\mu_1}  S_F (x-y) \gamma^{\nu_1} S_F (y-z) \gamma^{\mu} +
\cdots ]_{ji},
\ea
where $k_2$ is the momentum of one of emitted
gluons, $S_F (x - y)$ is the Feynman propagator of c-quark, the
dots in the square bracket denotes other five terms. In the limit
of $m_c \to\infty$, a $c$- or $\bar c$-quark moves with a small
velocity $v$, this fact enables us to describe nonperturbative
effect related to $\jpsi$ by NRQCD\cite{nrqcd}.
For the matrix element
$\langle 0 | \bar{c}_j (x_1) c_i (y_1) | \jpsi\rangle$
the expansion in $v$ can be performed with the result:
\beq
\label{nrqcdmatr}
\langle 0 |\bar{c}_j (x_1) c_i (y_1)| \jpsi
\rangle ~= - \frac{1}{6} ~(P_+ ~\gamma^{\ell} ~P_-)_{ij} ~\langle 0 |
\chi^{\dagger} \sigma^{\ell} \psi | \jpsi \rangle ~ e^{- i p\cdot (x_1
+ y_1)} + {\rm O} (v^2),
\eeq
where $\chi^{\dagger} (\psi)$ is the NRQCD field for $\bar{c} (c)$
quark, $\sigma^{\ell} (\ell = 1, 2, 3)$ is the Pauli matrix, and
\ba
P_{\pm} &=& \frac{1}{2} ~( 1 \pm \gamma^0 ) \nonumber \\
p &=& ( m_c, 0, 0, 0 ).
\ea
The matrix $\langle 0 | \chi^{\dagger}
\sigma^{\ell} \psi | \jpsi \rangle $ is proportional to the
polarization vector $\varepsilon ^{\ell} (\jpsi)$  at the considered
order. In this paper, we neglect the contribution from higher
orders in $v$, the momentum of $\jpsi$ is then approximated by $2p$.
It should be noted that effects at higher order of $v$
can be added with the expansion in  (\ref{nrqcdmatr}).


\begin{figure}[hbt]
\centering
\includegraphics[width=5cm,height=3cm]{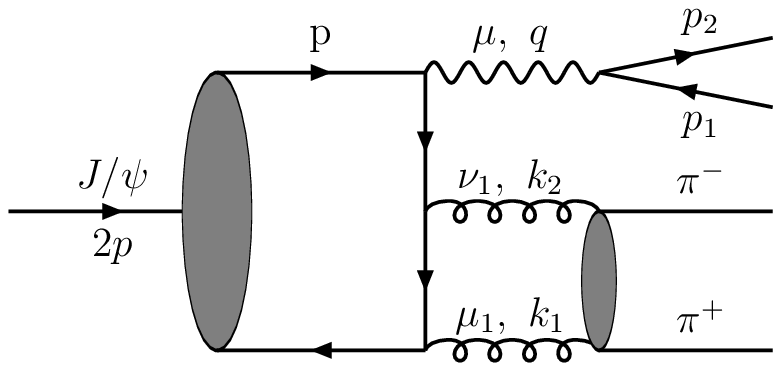}
\caption{ One of the Feynman diagrams for the exclusive decay of
$\jpsi$ into lepton pair and two pions.
\label{Feynman-dg1} }
\end{figure}

\par
Using  (\ref{nrqcdmatr}) we can write the S-matrix element as
\ba
\label{smatrix}
\smatr &=& \frac{- i}{24} Q_c e^2 ~g_s^2
~(2\pi)^4 ~\delta^{4} (2 p - k - q) L_{\mu} \cdot \frac{1}{q^2}
~\langle 0 | \chi^{\dagger} \sigma^{\ell} \psi | \jpsi \rangle
\nonumber\\
&& \times \int \frac{d^4 k_1}{(2 \pi)^4}  H^{\ell \mu \mu_1 \nu_1} (p, k, k_1)
~\Gamma_{\mu_1 \nu_1} (k, k_1) ,
\ea
and
\beq
\label{gammunu-d}
\Gamma^{\mu \nu} (k, k_1)
= \int d^4 x ~e^{- i k_1 x} ~\langle \twopi |
G^{a,\mu} (x) G^{a,\nu} (0)| 0 \rangle,
\eeq
where
$H^{\ell \mu \mu_1\nu_1} (p, k, k_1)$ is the amplitude for a $c\bar c$ pair
emitting a virtual photon and two gluons,
and this can be calculated with perturbative QCD.
The contributions in (\ref{smatrix}) can be
represented by Feynman diagrams. One of them is given in
Fig.\ref{Feynman-dg1}, where the kinematic variables are also
indicated. The nonperturbative object $\Gamma^{\mu \nu}(k, k_1)$
describes how two gluons are converted into the two pions.

\par
If the two pion system have a large momentum and a small invariant
mass, a twist expansion for the nonperturbative object
$\Gamma^{\mu \nu}(k, k_1)$ can be performed.
For convenience we will work
in the light-cone coordinate system, in which the components of $k$ are
given by

\beq
 k^\mu =(k^+, k^-,{\bf 0}), \ \ k^+ =(k^0+k^3)/\sqrt{2},\ \
 k^- =(k^0-k^3)/\sqrt{2}.
\eeq
In the light-cone coordinate system we introduce two light cone vectors
and a tensor:
\ba
n^{\mu} &=& (0, 1, 0, 0),  \ \
\tilde{n}^{\mu} = (1, 0, 0, 0),\nonumber \\
d_T^{\mu\nu} &=& g^{\mu\nu} -n^\mu \tilde{n}^\nu -n^\nu \tilde{n}^\mu,
\ea
and we take the gauge
\beq
 n\cdot G(x) =0 .
\eeq

\begin{figure}[htb]
\centering
\includegraphics[width=8cm,height=8cm]{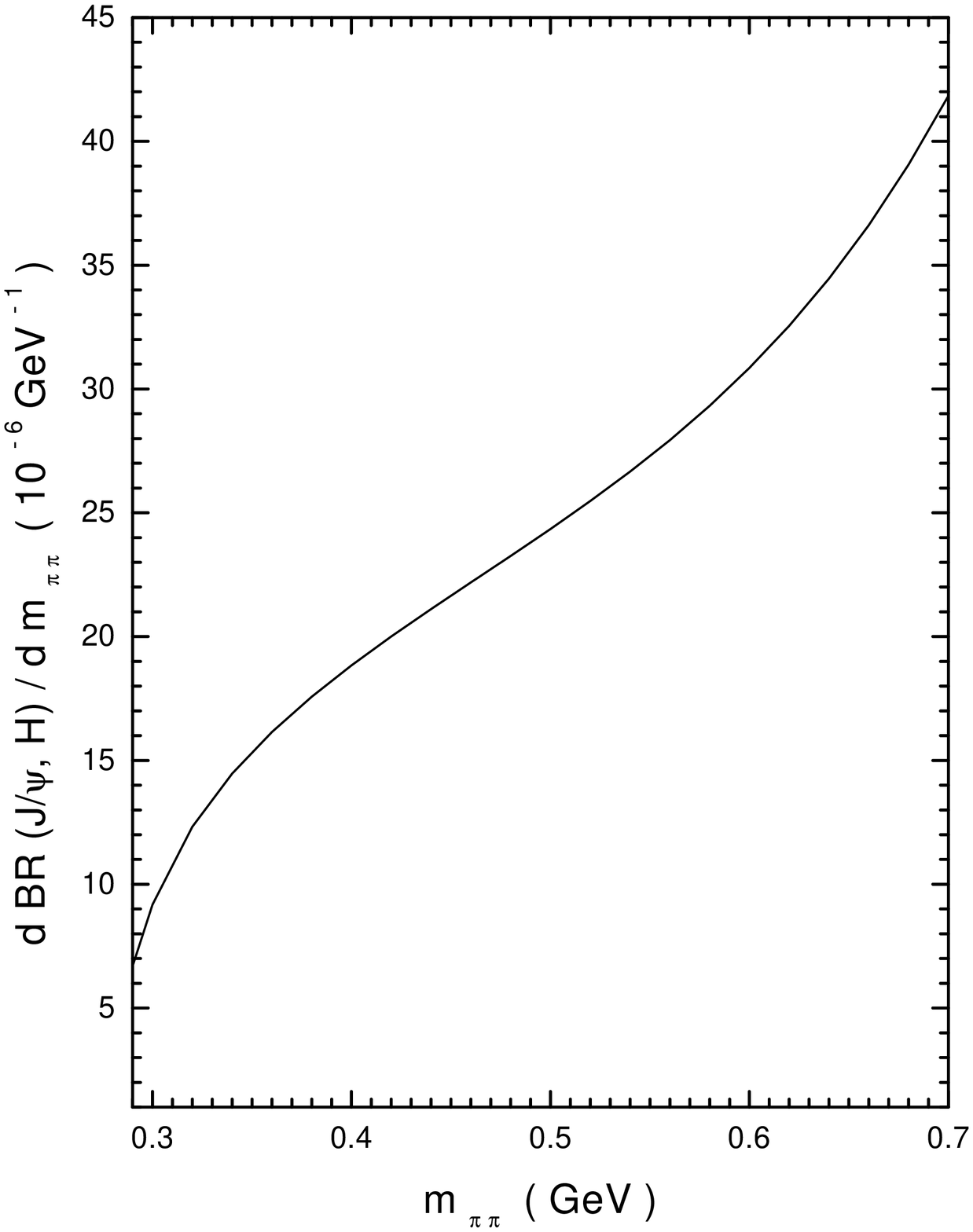}
\vspace*{-15mm}
\caption{The differential decay branching ratio of $\jpsi$,
$d ~{\rm BR} (\jpsi, H)/dm_{\pi\pi}$ as a function of $m_{\pi\pi}$
in unit of $ 10^{-6} {\rm GeV}^{-1}$ with the cuts given in the text.
\label{mpipi-ds}}
\end{figure}

\par
The $x$-dependence of the matrix element in (\ref{gammunu-d}) is
controlled by different scales.
The $x^-$-dependence is controlled by $k^+$, which is large in the
kinematic region we considered, while
the $x^+-$ and ${\bf x_T}$-dependence are controlled by the scale $k^-$
and $\Lambda_{QCD}$, which are small in comparison with $k^+$. With this
observation we can expand the matrix element in $x^+$ and in ${\bf x_T}$.
The resulted twist expansion of the Fourier transferred
matrix element $\Gamma^{\mu \nu} (k, k_1)$ is a collinear
expansion in $k^{-}$ and $k_T \sim \Lambda_{QCD}$.
Hence the expansion parameters of
$\Gamma^{\mu \nu} (k, k_1)$ are $k^- / k^+ $ and
$\Lambda_{QCD} / k^+ $, with $k^- / k^+ \leq 0.10 $ and
$\Lambda_{QCD} / k^+ \approx 0.20$ for $\jpsi$ in the
kinematic region considered.
At the leading order only twist-2 operators contributes to the matrix
element. We will neglect higher orders in the expansion, i.e., we only
keep contributions of twist-2 operators.
Then we obtain:

\beq
\label{gammunu}
\Gamma^{\mu \nu} (k, k_1) = (2 \pi)^4 ~\delta ( k_1^- )
~\delta^2 ( k_{1T} )~\frac{1}{k^+} ~\frac{1}{x_1 (1 - x_1)} \nonumber \\
\left[ \frac{1}{2} ~d_T^{\mu \nu}
~\Phi^{G} ( x_1, \zeta, m_{\pi \pi} )  \right] ,
\eeq
with
\ba
\label{phig}
\Phi^{G} ( x_1, \zeta, m_{\pi \pi} )&=& \frac{1}{k^+}
~\int \frac{d x^-}{2 \pi} ~e^{ - i k_1^+ x^-}
\times \langle \twopi |G^{a,+\mu} (x^-n) G_{\mu}^{a,+} (0)| 0 \rangle,
\nonumber \\
x_1 &=& \frac{k_1\cdot n}{k\cdot n}, \ \ \
\zeta = \frac{k_{\pi^+} \cdot n}{k \cdot n}.
\ea
$\Phi^{G} ( x_1, \zeta, m_{\pi \pi} )$ is the gluonic distribution
amplitude which describe how a pion pair with helicity $\lambda = 0$
is produced by two
collinear gluons; this represents a nonperturbative effect and can
only be calculated with nonperturbative methods or extracted from
experiment. As it stands, it is gauge invariant in the gauge
$n\cdot G(x) =0$. In other gauges we need to supply a Wilson line
operator to make it gauge invariant.
With (\ref{gammunu}) it is
straightforward to obtain the $S$-matrix element at the tree-level
in our approximation:
\ba
\label{smatrix-f}
\smatr &=& \frac{- i}{24} Q_c e^2 g_s^2 (2 \pi)^4 ~\delta^{4}
(2 p - k - q) L^{\mu} \cdot \frac{1}{q^2}
\langle 0 | \chi^{\dagger} \sigma^\ell \psi | \jpsi \rangle  \nonumber \\
&& \times \int_0^1  d x_1 \frac{\Phi^G ( x_1, \zeta, m_{\pi \pi} )}
{x_1 (1 - x_1) }\cdot \left[ \frac{1}{2} d_{T}^{\mu_1 \nu_1} \cdot
H_{\ell \mu \mu_1 \nu_1} (p, k, k_1) \right],
\ea
with
\beq
\label{halfdh}
\frac{1}{2} d_{T}^{\mu_1 \nu_1} \cdot H_{\ell \mu \mu_1 \nu_1} (p, k, k_1)
= \frac{16}{M_{\psi}^{2}} ~\tilde{n}_{\mu} n_{\ell} -
\frac{16}{M_{\psi}^{2} - q^2} ~g_{\mu \ell}~,
\eeq
where $M_{\psi}$ is the mass of $\jpsi$.
In (\ref{halfdh}) we have neglected  the mass $m_{\pi\pi}$,
since the effect of $m_{\pi\pi}$ should be combined with
effects of twist-4 operators as a correction to the above result.
In (\ref{smatrix-f}) the nonperturbative effect related to
$\jpsi$ and that to the two-pion system are separated, the former
is represented by a NRQCD matrix element, while the later is
represented by the distribution amplitude of two gluons in
the isoscalar pion pair, which is defined in (\ref{phig}) in the
gauge $ n \cdot G (x) = 0$.

\begin{figure}[htb]
\centering
\includegraphics[width=8cm,height=8cm]{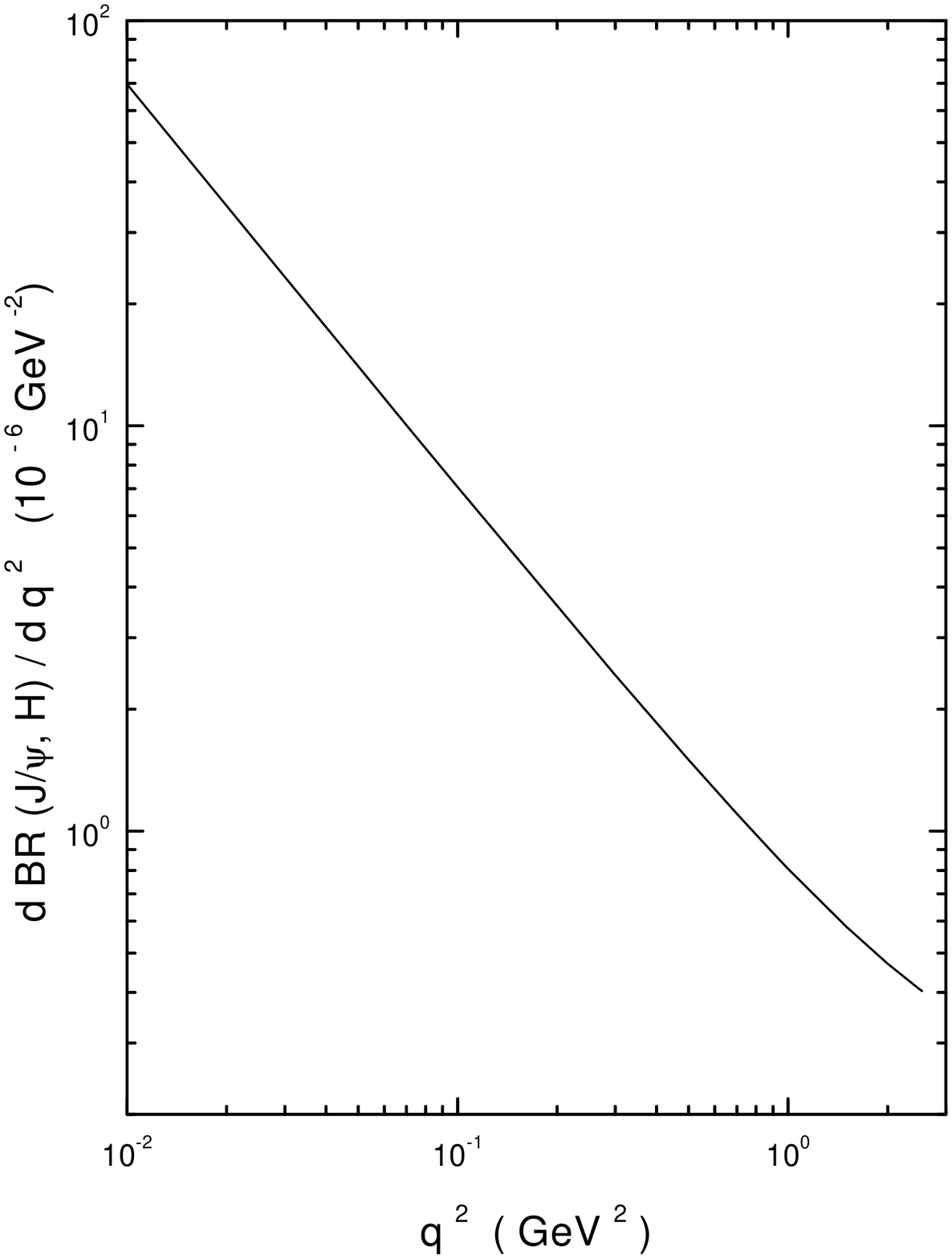}
\vspace*{-15mm}
\caption{The differential decay branching ratio of $\jpsi$,
$d ~{\rm BR} (\jpsi, H)/d q^2$ as a function of $ q^2$
in unit of $ 10^{-6} {\rm GeV}^{-2}$ with the cuts given in text.
\label{qsq-ds}}
\end{figure}

\par
The kinematics of the decay can be fully described by five
variables as in $K_{e 4}$ decay\cite{cabibbo}:
\begin{enumerate}
\item
$m_{\pi \pi}^{2}$, the invariant mass squared of the pion pair;

\item
$q^2 = (p_1 + p_2)^2$,
the invariant mass squared of the lepton pair;

\item $\theta_{\pi}$, the polar angle of the $\pi^+$ in the rest frame of
the pion pair with respect to the moving direction of the
pion pair in the $\jpsi$ rest frame;

\item
$\theta_{l}$, the polar angle of the $l^+$ in the rest frame of
lepton pair with respect of the moving direction of the lepton pair
in the $\jpsi$ rest frame;

\item $\phi$, the azimuthal angle
between the two planes in which the pion pair and the lepton pair
lies respectively.
\end{enumerate}
In terms of these variables, the differential decay width can be written as
\beq
\label{dgam}
d \Gamma = \frac{1}{(2 \pi)^8} \cdot \frac {\pi^2}{32} \cdot
\frac{|\vec{k}|}{M_{\psi}} \cdot \beta ~\beta_l
~\overline{\sum} |M|^2 d q^2 d m_{\pi \pi}^{2}
d \cos \theta_{\pi} d \cos \theta_l d \phi ,
\eeq
where $\beta$ and $\beta_l$ are defined as:
\beq
\beta = \sqrt{1 -
\frac{4 ~m_{\pi}^{2} }{m_{\pi \pi}^{2}}} ~,
~~~~\beta_l  = \sqrt{1 -
\frac{4 ~m_{l}^{2} }{q^2}} ~,
\eeq
$\overline{\sum} |M|^2$ is the absolute squared matrix element of
the decay, summed over final state spins and averaged over initial state
spin. From  (\ref{smatrix-f}) and  (\ref{halfdh}), we have
\ba
\label{msq}
\overline{\sum} |M|^2 &=& \frac{1}{24^2} Q_c^2 e^4 g_s^4 \cdot
  \frac{1}{q^4} \cdot
  |\langle 0 | \chi^{\dagger} \bfsig \psi | \jpsi \rangle|^2
\nonumber \\
&& \times \frac{512 q^2 [(M_{\psi}^{2} + q^2 ) + (M_{\psi}^{2} - q^2)
   \cos ^2 \theta_l ]}{3 M_{\psi}^{2} (M_{\psi}^{2} - q^2)^2}
\nonumber \\
&& \times \left| \int_0^1 d x_1 \frac{\Phi^G ( x_1, \zeta, m_{\pi \pi} )}
   {x_1 (1 - x_1) } \right|^2 ,
\ea
which is independent of the azimuthal angle $\phi$, the spin average for
$\jpsi$ is implied in the squared matrix element.

\begin{figure}[htb]
\centering
\includegraphics[width=8cm,height=8cm]{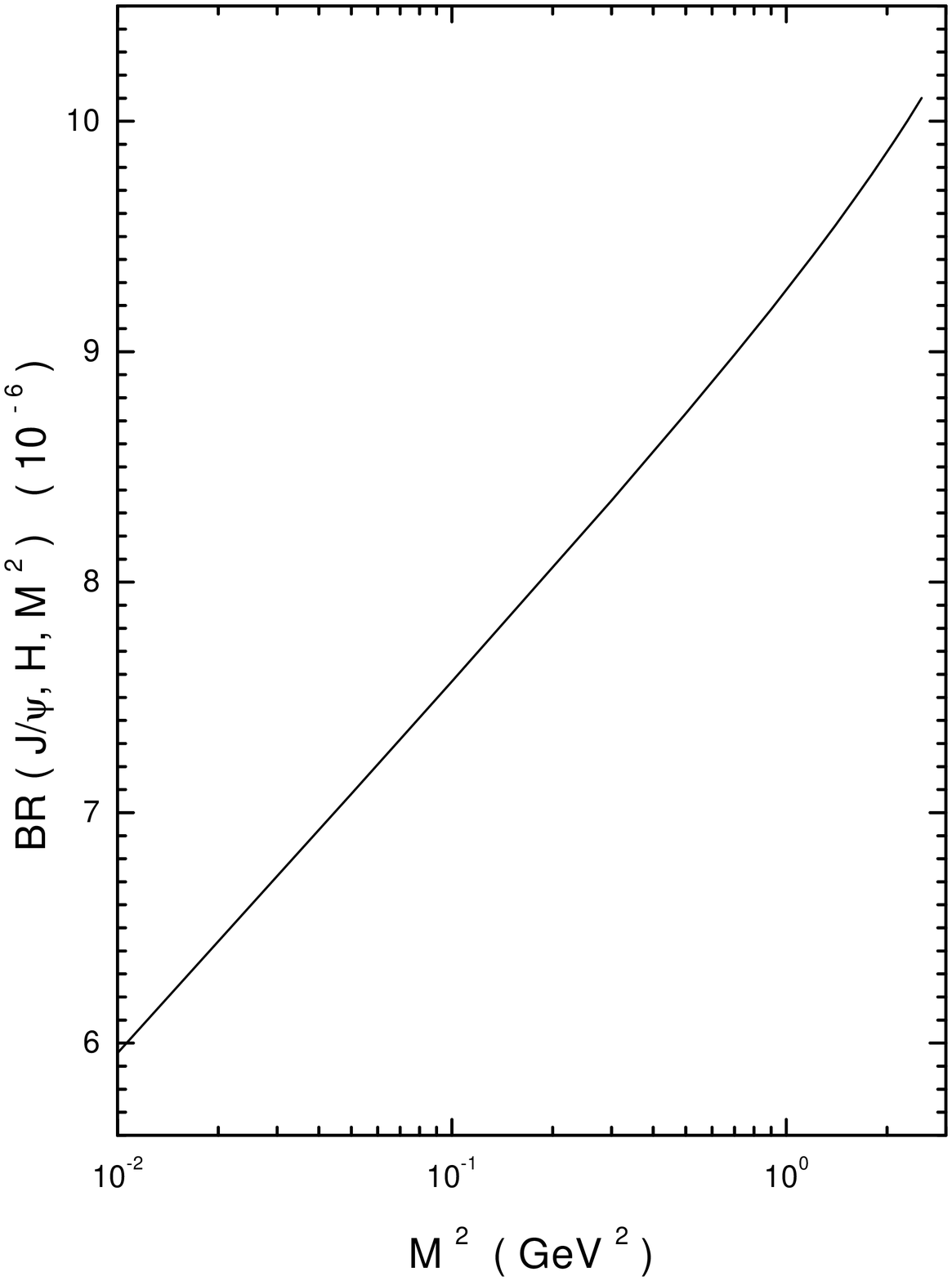}
\vspace*{-10mm}
\caption{The decay branching ratio of $\jpsi$,
${\rm BR} (\jpsi, H, M^2)$ as a function of $M^2$ with
$M^2 > q^2$
in unit of $10^{-6}$. The other cuts are the same.
\label{qsqup-ds}}
\end{figure}

\par
To give numerical predictions, the nonperturbative
inputs, the NRQCD matrix element and the distribution amplitude
of two gluons in the isoscalar pion pair, are needed.
The NRQCD matrix element is related to the
wave-function of $\jpsi$ in potential models and can be estimated
with these models. It can also be calculated with lattice QCD or
extracted from experiment.
In this paper, we use leptonic decay of $\jpsi$ to determine
the NRQCD matrix element, i.e.,

\beq
\Gamma ( \jpsi \to e^+ e^-) = \frac{2 \pi Q_c^2 \alpha^2} {3 m_c^2}
\cdot
|\langle 0 | \chi^{\dagger} \bfsig \psi | \jpsi \rangle|^2.
\eeq

\begin{figure}[htb]
\centering
\includegraphics[width=8cm,height=8cm]{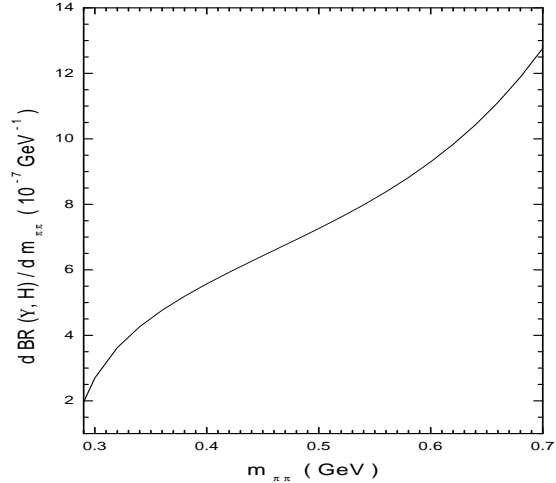}
\vspace*{-15mm}
\caption{The differential decay branching ratio of $\Upsilon (1S)$,
$d ~{\rm BR} (\Upsilon, H)/dm_{\pi\pi}$ as a function of $m_{\pi\pi}$
in unit of $ 10^{-7} {\rm GeV}^{-1}$ with the cuts given in the text.
\label{bmpipi-ds}}
\end{figure}

\noindent
The distribution amplitude $\Phi^{G}( x_1, \zeta, m_{\pi \pi} )$
is not determined with experiment, a detailed study
of $\Phi^{G}(x_1,\zeta,m_{\pi \pi})$ can be found in \cite{D2,KMP,L1}.
For our numerical prediction we will use their results for $\Phi^{G}
( x_1, \zeta, m_{\pi \pi} )$, in which asymptotic form of
$\Phi^{G} ( x_1, \zeta, m_{\pi \pi} )$ is taken as an Ansatz for
$\Phi^{G} ( x_1, \zeta, m_{\pi \pi} )$. It should be noted that
the renormalization scale $\mu$ should be taken as $ M_{\psi} $ in
our case. Because it is
not very large, the actual shape of $\Phi^{G} ( x_1, \zeta, m_{\pi
\pi} )$ may look dramatically different than that of the
asymptotic form. Keeping this in mind we take the form
$\Phi^{G} ( x_1, \zeta, m_{\pi \pi} )$ as that given in \cite{L1}:
\beq
\Phi^{G} ( x_1, \zeta, m_{\pi \pi} )= - 60 ~M_2^G ~x_1^2 ~(1 - x_1)^2
\left[ \frac{3 C - \beta^2}{12} ~
f_0 ( m_{\pi\pi}) ~ P_0 (\cos \theta_{\pi}) \right. \nonumber \\
\left. - \frac{\beta^2}{6} ~f_2 (m_{\pi\pi}) ~P_2 (\cos \theta_{\pi})
\right] ~,
\eeq
where $\zeta$ is related to $\theta_{\pi}$ and $m_{\pi\pi}$ by
\beq
\beta ~\cos \theta_{\pi}  = 2 ~\zeta - 1,
\eeq
$C$ is a constant and takes $C = 1 + b ~m_{\pi}^{2} + O (m_{\pi}^{4}) $
with $b \simeq - 1.7 {\rm GeV}^{- 2}$ \cite{KMP,L1}, $M_2^G$ is determined
by gluon fragmentation into a single pion,
its asymptotic
value  is
\beq
\label{m2g}
M_2^G = \frac{ 4 ~C_F}{ N_f + 4 ~C_F}.
\eeq
$f_0 (m_{\pi\pi})$ and $f_2 (m_{\pi\pi})$  are the Omn\`{e}s
functions for ${\rm I} = 0$ s- and d-wave $ \pi \pi$ scattering,
respectively. The Omn\`{e}s function  $f_2 (m_{\pi\pi})$ is
dominated by the $f_2 (1270)$ resonance resulting a peak at
$m_{\pi\pi} = 1.275 {\rm GeV}$, while the Omn\`{e}s function  $f_0
(m_{\pi\pi})$ in the relevant $m_{\pi\pi}$ region we studied
($m_{\pi\pi} \leq 0.70 {\rm GeV}$) can be calculated by the chiral
perturbative theory, the result is \cite{donoghue}

\begin{equation}
f_0(m_{\pi\pi})= 1 + \frac{m_{\pi\pi}^{2}}{192 \pi^2 f_{\pi}^{2}}
 + \frac{2 m_{\pi\pi}^{2} - m_{\pi}^{2}}{32 \pi^2 f_{\pi}^{2}}
\left[ \beta \ln \left(\frac{1 - \beta}{1 + \beta}
\right) + 2 + i \pi \beta \right] ,
\end{equation}
where $f_{\pi}=93~{\rm MeV}$ ~is the pion decay constant.

\begin{figure}[htb]
\centering
\includegraphics[width=8cm,height=8cm]{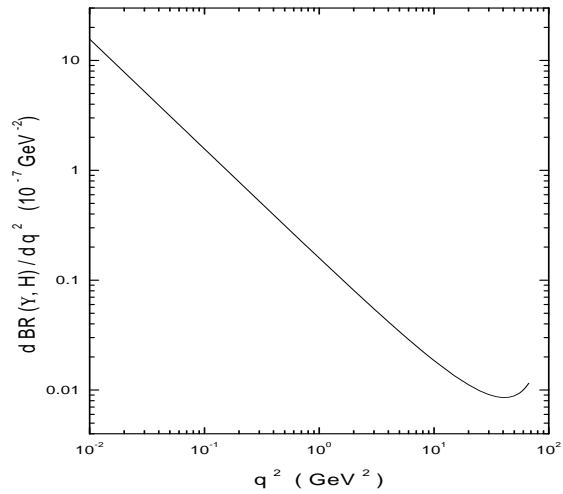}
\vspace*{-15mm}
\caption{The differential decay branching ratio of $\Upsilon (1S)$,
$d ~{\rm BR} (\Upsilon, H)/d q^2$ as a function of $ q^2$
in unit of $ 10^{-7} {\rm GeV}^{-2}$ with the cuts given in the text.
\label{bqsq-ds}}
\end{figure}

\par
Cuts must be used to select the kinematic region where
the two pion system has a large momentum and a small invariant mass.
We use the cuts: $k^+ \geq 10 ~k^-$, $k^0 + \vert{\bf k}\vert \geq 2.0
{\rm GeV}$ and $2 m_{\pi} \leq m_{\pi \pi} \leq 0.70 {\rm GeV}$,
this corresponds to $q^2 \leq 2.5 {\rm GeV}^2$ for $\jpsi$
and $q^2 \leq 67 {GeV}^2$ for $\Upsilon (1S)$.

\par
With these results we are able to predict the  decay branching
ratio in the considered region. The quark masses are take
as $m_c = \frac{1}{2} M_{\psi}$ and $m_b = \frac{1}{2} M_{\Upsilon}$.
$\alpha_s (2 m_c) = 0.31$ for $\jpsi$, $\alpha_s ( 2 m_b) = 0.21$
for $\Upsilon (1S)$. All other parameters needed are taken from
\cite{pdg2000}. The $m_{\pi \pi}$ distribution of $\jpsi$ decay,
integrated over $|\cos \theta_{\pi}|
\leq 1.0, |\cos \theta_l| \leq 1.0, 0 \leq \phi \leq 2 \pi$, and
$4 m_l^2 \leq q^2 \leq 2.5 {\rm GeV}^2$, denoted as
$d ~{\rm BR} (\jpsi, H)/dm_{\pi\pi}$, is shown in Fig. \ref{mpipi-ds}.
The $q^2$ distribution of $\jpsi$ decay,
integrated over $|\cos \theta_{\pi}|
\leq 1.0, |\cos \theta_l| \leq 1.0, 0 \leq \phi \leq 2 \pi$, and
$2 m_{\pi} \leq m_{\pi \pi} \leq 0.70  {\rm GeV}$, denoted as
$d ~{\rm BR} (\jpsi, H)/d q^2$, is shown in Fig. \ref{qsq-ds}.
The $q^2$ distribution decreases rapidly as $q^2$ increases, this is mainly
due to the $q^{-2}$ factor of the photon propagator.  This behavior
is shown in another way in Fig. \ref{qsqup-ds}, where the decay branching
ratio as a function of the cut $M^2$ with $q^2<M^2$.
We see from this figure, for $M^2 = 10^{-2}, 10^{-1}, 10^0 {\rm GeV}^2$,
the corresponding contributions are $59 \%, 75 \%, 92 \%$ to
the branching ratio in the whole region considered, respectively.
Integrating over the kinematic region we considered, the
decay branching ratio is $1.0 \times 10^{-5}$ for $\jpsi$, among which
the s-wave and d-wave contributions are $9.2 \times 10^{-6}$ and
$8.6 \times 10^{-7}$ respectively, i.e., the d-wave contribution is
suppressed in the kinematic region here.
The results indicate that this decay mode as well as
the distributions of $m_{\pi \pi}$ and $q^2$ can be observed at
BES II, and at the proposed BES III and CLEO-C.
In above numerical calculations, the renormalization scale
of the effective QCD coupling is taken to be $\jpsi$ mass
with $\Lambda^{(4)}_{QCD} = 280 {\rm MeV}$[14], i.e.,
$\alpha_s (2 m_c) = 0.31$. If this scale is taken to be
$m_c$, the corresponding decay branching ratio of
$\jpsi$ in the considered kinematic region increases
by a factor of $2.1$ by using  (18). Hence,
the decay branching ratio is conservatively predicted.

\begin{figure}[htb]
\centering
\includegraphics[width=8cm,height=8cm]{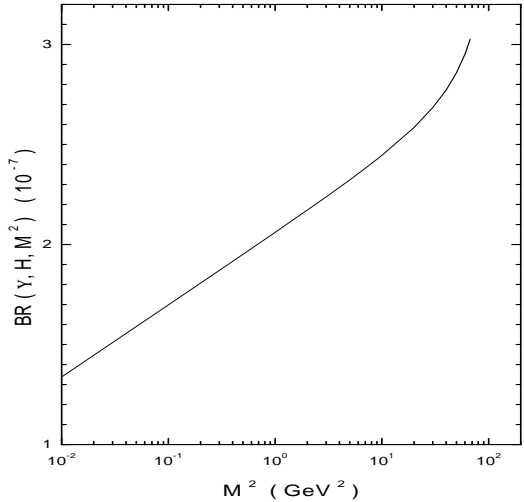}
\vspace*{-12mm}
\caption{The decay branching ratio of $\Upsilon (1S)$,
${\rm BR} (\Upsilon, H, M^2)$ as a function of $M^2$
with $M^2 >q^2$
in unit of $10^{-7}$. The other cuts are the same.
\label{bqsqup-ds}}
\end{figure}

\par
The corresponding differential decay branching ratios for $\Upsilon (1S)$
decay are shown in Figs. \ref{bmpipi-ds}-\ref{bqsqup-ds}.
The kinematic region
we studied for $\Upsilon (1S)$ decay is $|\cos \theta_{\pi}| \leq 1.0,
|\cos \theta_l| \leq 1.0, 0 \leq \phi \leq 2 \pi,
2 m_{\pi} \leq m_{\pi \pi} \leq 0.70  {\rm GeV}$,
and $ 4 m_{l}^{2} \leq q^2 \leq 67 {\rm GeV}^2$.
It is interesting to observe that in Fig. \ref{bqsq-ds} there is a
turn-over near $q^2 = 40 {\rm GeV}^2$.
Similar to $\jpsi$ case, the dominant contribution to the decay
of $\Upsilon (1S)$ comes also from small $q^2$ region.
For $M^2 = 10^{-2}, 10^{-1}, 10^0 $, and $ 10^1 {\rm GeV}^2$,
the corresponding contributions are $44 \%, 56 \%, 68 \% $, and $ 81 \% $
to the branching ratio in the whole region considered, respectively.
Integrating over the kinematic region we considered, the
decay branching ratio is $3.0 \times 10^{-7}$, among which
the s-wave and d-wave contributions are $2.8 \times 10^{-7}$ and
$2.6 \times 10^{-8}$ respectively, i.e., the d-wave contribution is
also suppressed in the kinematic region here. For the decay of
$\Upsilon$ one may allow $m_{\pi\pi}$ to be larger than that
in the decay of $\jpsi$, i.e., 0.7GeV, because the phase space is large.
With a large upper cut for $m_{\pi\pi}$, the branching ratio can
become large. But $f_0(m_{\pi\pi})$ determined
with chiral perturbation theory may become unreliable for large
$m_{\pi\pi}$.

\par
It should be noted that the two-pion state is produced
with the helicity $\lambda =0$ at the level of leading
twist.
It can be a mixture of states with different angular momenta
$L$. Because of parity conservation, isospin symmetry and
Bose-Einstein statistics, $L$ can only be even. Our numerical results
show that the state is mainly in a s-wave state.
At levels of higher twist it is possible that the two-pion
state is produced with $\lambda\neq 0$. Following
the analysis for the radiative decay of $\Upsilon$ into
$f_2(1270)$\cite{F2}, one can show that the two-pion
state can be produced with $\vert\lambda\vert =1$ and $2$
at order of twist 3 and of twist 4, respectively.


\section{Leptonic decays of $\jpsi$ and $\Upsilon (1S)$ combined with
         two soft pions}
In this section, we study the leptonic decays of
$\jpsi$ and $\Upsilon (1S)$ combined with two soft pions.
In contrast to the decays studied in the last section
the gluons, which are emitted by the heavy quarks in the
quarkonium and are converted into the pions, are soft.
The emission of soft gluons can be studied by employing
an expansion in the inverse of the heavy quark mass $m_Q$.
It is shown in \cite{jpma} that at leading order
of the expansion the decay amplitude in this kinematic region can
be factorized into three parts: the first part is a NRQCD matrix element
representing the bound-state effect of heavy quarkonium, the second part
is a matrix element of a correlator of electric chromofields, which
indicates the nonperturbative effect of the soft gluons converted into
the soft pion pair, the third part consists of some coefficients.
It should be emphasized that the results can be derived
without using perturbative QCD.
In this section
we present a model for the matrix element of the correlator of electric
chromofields, and give numerical predictions.
The S-matrix for the $\jpsi$ decay is\cite{jpma}
\ba
\label{smatrix-s}
\smatr  &=& i \frac{2}{3} Q_c e^2  (2 \pi)^4 ~\delta^{4}
(2 p - k - q) L_{\mu} \cdot \frac{ g^{\mu \ell}}{q^2}
\langle 0 | \chi^{\dagger} \sigma^\ell \psi | \jpsi \rangle  \nonumber \\
&& \times \frac{1}{m_c} \cdot \frac{1}{(k^0)^2} \cdot T_{\pi \pi} (k)
+ {\cal O} (\frac{1}{m_c^2}) + {\cal O} (v^2), \nonumber \\
T_{\pi \pi} ( k ) &=& \int d \tau  \frac{1}{1 + \tau - i 0^+} \cdot
\frac{1}{1 - \tau - i 0^+} h (\tau,k),
\ea
where the momenta are denoted as the same in the last section.
For soft pions we have $\vert{\bf k}\vert\ll m_Q$ and $m_{\pi\pi}\ll m_Q$.
$h (\tau, k)$ is the distribution amplitude for the soft gluons converted
into two soft pions. It is defined as
\beq
\label{htauk}
h (\tau,k) = \frac{g_s^2}{2 \pi} \int_{- \infty}^{ + \infty}
d t e^{i \tau k^0 t} \langle \pi^+ \pi^- |{\bf E}^a (t,{\bf 0}) \cdot
\left[ P~ {\rm exp} \left\{ - i g_s \int_{-t}^{t} d x^0 G^{0,c} (x^0, {\bf 0})
\tau^c \right\} \right] _{ab} {\bf E}^b (-t,{\bf 0}) | 0 \rangle,
\eeq
where $P$ denotes path-ordering and $\tau^c$ is the generator of SU(3) in adjoint
representation, $(\tau^c)_{ab} = - i f_{abc}$.
Because of the energy conservation $h (\tau,k) = 0$ if $|\tau| > 1$. The term with
$g^{\mu \ell}$ in  (\ref{smatrix-s}) is expected in the heavy quark limit. In
this limit emitted gluons will not change the spin of the heavy quarks,
hence the spin of $\jpsi$ is transferred to the virtual photon. Therefor the
helicity of the two-pion state is zero.

\par
The function $h (\tau,k)$ is unknown. We make an Ansatz for the
$\tau$-dependence in the function, this Ansatz is motivated by the results
used in the last section. We assume
\beq
\label{htauk-ak}
h (\tau,k) = a (k) (1 - \tau)^2 (1 + \tau)^2,
\eeq
the function $a(k)$ can be obtained by integrating $h (\tau,k)$ over $\tau$,
we obtain:
\beq
a (k) = \frac{15 \pi}{4 k^0} \langle \pi^+ \pi^- | \alpha_s {\bf E}^a (0) \cdot
{\bf E}^a (0) | 0 \rangle ,
\eeq
hence for $|\tau| \leq 1$,
\beq
\label{htauk-f}
h (\tau, k) = \frac{15 \pi}{4 k^0} (1 - \tau)^2 (1 + \tau)^2 \cdot
 \langle \pi^+ \pi^- | \alpha_s {\bf E}^a (0) \cdot
{\bf E}^a (0) | 0 \rangle.
\eeq
The matrix element of local chromoelectric fields
$\langle \pi^+ \pi^- | \alpha_s {\bf E}^a (0) \cdot
{\bf E}^a (0) | 0 \rangle $, which appears in the decay
amplitude of $\Psi^{\prime} \to \jpsi \pi^+ \pi^-$ in
the  QCD multipole expansion method\cite{voloshin,pe,novikov,yan},
can be written in our notations as \cite{novikov}
\ba
\label{mt-ee}
\langle \pi^+ \pi^- | \alpha_s {\bf E}^a (0) \cdot {\bf E}^a (0) | 0 \rangle
&=& \frac{2 \pi}{9} \langle \pi^+ \pi^- | \theta_{\mu}^{\mu} | 0 \rangle
+ \langle \pi^+ \pi^- | \alpha_s (\mu) \theta^{G}_{00} (\mu) | 0 \rangle
\nonumber \\
&=& \frac{2 \pi}{9} \langle \pi^+ \pi^- | \theta_{\mu}^{\mu} | 0 \rangle
 - \frac{1}{3} \alpha_s (\mu) M_2^G (\mu) (k^0)^2
 \left( 1 + \frac{2 m_{\pi}^{2}}{m_{\pi\pi}^{2}} \right) P_0 (\cos \theta_{\pi})
\nonumber \\
&&  + \frac{1}{3} \alpha_s (\mu) M_2^G (\mu) |{\bf k}|^2 \beta^2
P_2 (\cos \theta_{\pi}),
\ea
where $\theta_{\mu \nu}$ is the total energy-momentum tensor of QCD,
$\theta_{\mu \nu}^{G}$ is the gluonic component of it,
$M_2^G (\mu)$ is determined by gluon fragmentation into one
pion as before. In \cite{novikov}, including ${\cal O} (m_{\pi}^{2})$
corrections,
$\langle \pi^+ \pi^- | \theta_{\mu}^{\mu} | 0 \rangle = q^2 + 2 m_{\pi}^{2}$
is obtained from some general considerations.  This coincides with
the result of chiral perturbation theory at leading order of chiral expansion.
Since the kinematic region we considered is only part of the whole phase
space and $m_{\pi \pi}$ is not very near $\pi^+ \pi^- $ threshold, we expect
that the correction from
next-to-leading order of chiral perturbation theory to be important, so we use
the expression derived from  chiral perturbation theory at next-to-leading order
for $\langle \pi^+ \pi^- | \theta_{\mu}^{\mu} | 0 \rangle $,
i.e. \cite{donoghue},
\beq
\langle \pi^+ \pi^- | \theta_{\mu}^{\mu} | 0 \rangle
= (m_{\pi \pi}^{2}  + 2 m_{\pi}^{2}) f_0 (m_{\pi \pi}) + b_{\theta} m_{\pi \pi}^{4}
\eeq
with $ b_{\theta} = 2.7 {\rm GeV}^{-2}$.

\begin{figure}[htb]
\centering
\includegraphics[width=8cm,height=8cm]{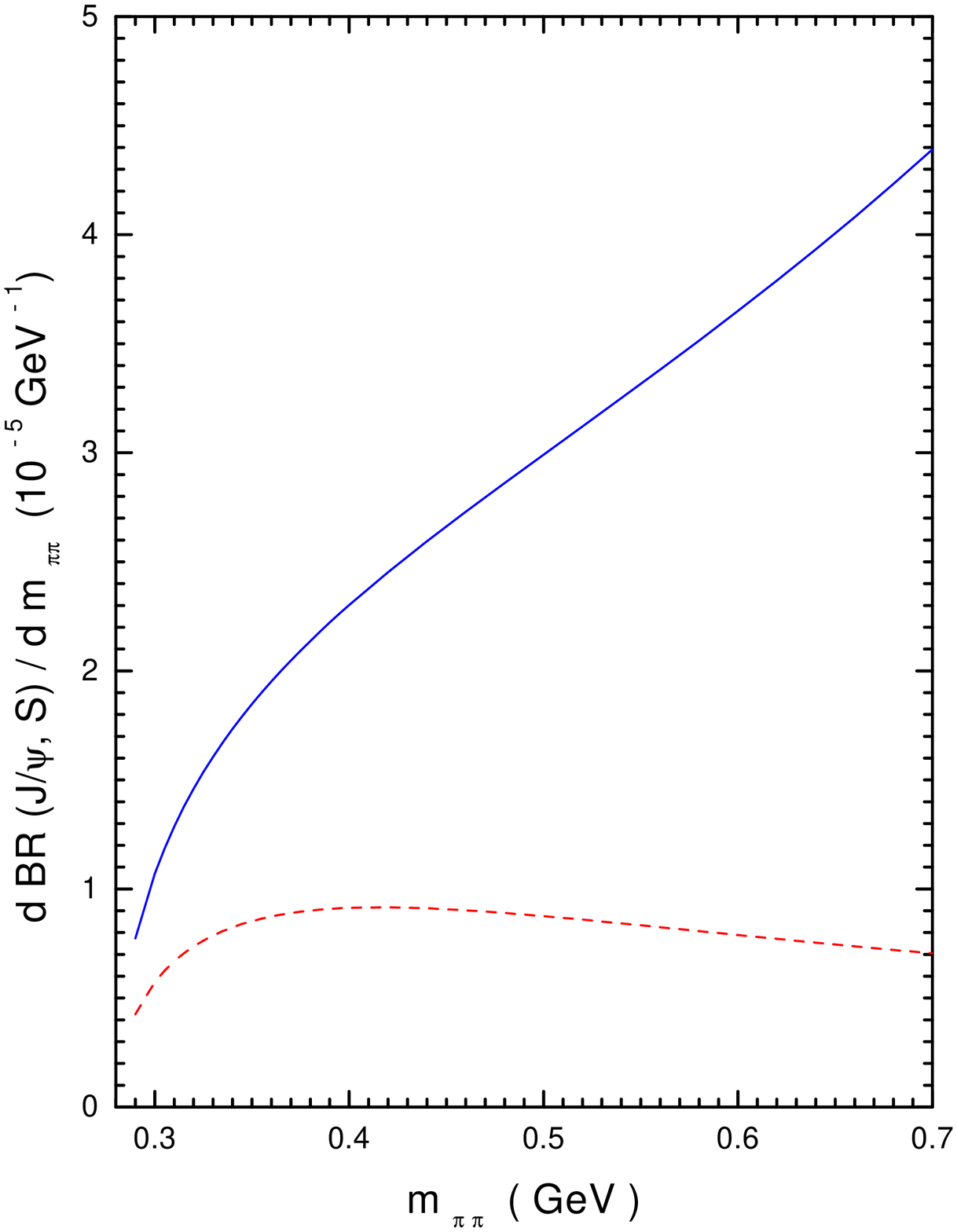}
\vspace*{-15mm}
\caption{The differential decay branching ratio of $\jpsi$,
$d ~{\rm BR} (\jpsi, S)/dm_{\pi\pi}$ as a function of $m_{\pi\pi}$
in unit of $ 10^{-5} {\rm GeV}^{-1}$ with the cuts.
The dashed line denotes the distribution by using the leading order result of
chiral perturbation theory for $\langle \pi^+ \pi^- | \theta_{\mu}^{\mu} | 0 \rangle $,
while the solid line denotes
the distribution by adding one-loop correction
to the matrix element.
\label{psism2pi}}
\end{figure}

\par
With these results, we are able to predict the shape of the differential decay
branching ratio numerically.  We use the cuts $0 \leq |{\bf k}| \leq \frac{1}{10} M_{\psi}$
and $2m_\pi < m_{\pi\pi} <0.7$GeV to make the pions to be soft.
In Fig. \ref{psism2pi}
The differential decay branching ratio of $\jpsi$,
$d ~{\rm BR} (\jpsi, S)/dm_{\pi\pi}$ as a function of $m_{\pi\pi}$
in unit of $ 10^{-5} {\rm GeV}^{-1}$ is shown. We use
$\alpha_s (\mu) =0.7$ and
$M_2^G (\mu)=0.5$ as used in \cite{novikov}.
The solid line denotes the distribution by using (30),  while the
dashed line denotes the distribution by using
$\langle \pi^+ \pi^- | \theta_{\mu}^{\mu} | 0 \rangle$ at the leading
order of chiral perturbation theory.
Integrating over $2 m_{\pi} \leq m_{\pi \pi} \leq 0.70 {\rm GeV}$,
the decay branching ratios for $\jpsi$ in the considered kinematic region
are $1.8 \times 10^{-5}$ and $3.5 \times 10^{-5}$, by using the result
at leading- and next-to-leading order of chiral perturbation theory respectively,
indicating the importance of the next-to-leading order
chiral corrections to the matrix
element of the QCD total energy-momentum tensor.
It should be noted all our numerical results are insensitive to
the values of $M_2^G (\mu)$ and $\alpha_s (\mu)$,
by varying the value of $M_2^G (\mu)$ from $0$ to its
asymptotic value  (\ref{m2g}), all numerical results are changed less than
$20 \%$. Since $\alpha_s (\mu)$ appears always with $M_2^G (\mu)$
in the form $ \alpha_s (\mu) \cdot M_2^G (\mu)$, the same is also true for
$\alpha_s (\mu)$. Our results indicate that this decay mode and
the $m_{\pi \pi}$ distribution
are observable at BES II and the proposed BES III and CLEO-C.
Experiment study of the decay can test our model for $h (\tau, k)$ or
extract it. This will provide information how gluons are converted
into two soft pions.

\par
Although
we have made the Ansatz for the function $h(\tau,k)$ in (26), where
the shape as a function of $\tau$ is fixed, and the parameter $a(k)$
is just a normalization factor determined by the matrix element
$\langle \pi^+ \pi^- | \alpha_s {\bf E}^a (0) \cdot {\bf E}^a (0) | 0 \rangle$,
but we can expect that our results for the
branching ratio will be not changed dramatically with a change of the shape,
because the normalization factor is fixed.

\begin{figure}[htb]
\centering
\includegraphics[width=8cm,height=8cm]{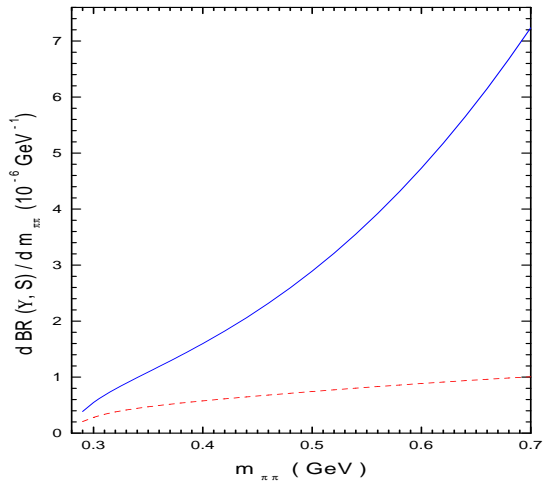}
\vspace*{-15mm}
\caption{The differential decay branching ratio of $\Upsilon (1S)$, referring
as $d ~{\rm BR} (\Upsilon, S)/dm_{\pi\pi}$, as a function of $m_{\pi\pi}$
in unit of $ 10^{-6} {\rm GeV}^{-1}$.
The dashed line denotes the distribution by using the leading order result of
chiral perturbation theory for $\langle \pi^+ \pi^- | \theta_{\mu}^{\mu} | 0 \rangle $,
while the solid line denotes
the distribution by adding one-loop correction
to the matrix element.
\label{bbsm2pi}}
\end{figure}

\par
The corresponding $m_{\pi \pi}$ distribution for $\Upsilon (1S)$, referring
as $d ~{\rm BR} (\Upsilon, S)/dm_{\pi\pi}$,
in unit of $ 10^{-6} {\rm GeV}^{-1}$ $0 \leq |{\bf k}| \leq \frac{1}{10} M_{\Upsilon}$
and $2m_\pi < m_{\pi\pi} <0.7$GeV is shown
in  Fig. \ref{bbsm2pi}.  The solid line denotes
the distribution using next-to-leading order chiral perturbative theory to determine
the matrix element $\langle \pi^+ \pi^- | \theta_{\mu}^{\mu} | 0 \rangle $, while the
dashed line denotes the distribution using leading order chiral perturbative theory
for this matrix element. Integrating over $2 m_{\pi} \leq m_{\pi \pi} \leq 0.70 {\rm GeV}$,
the decay branching ratios for $\Upsilon (1S)$ in the considered kinematic region
are $1.5 \times 10^{-6}$ and $3.5 \times 10^{-6}$,by using the result
at leading- and next-to-leading order of chiral perturbation theory respectively.
With the numerical results the decay mode may be difficult to be observed even at CLEO-C.
However, we can learn from comparing Figs. \ref{psism2pi} and \ref{bbsm2pi} that
when the phase space becomes larger, the next-to-leading order chiral corrections to
the matrix element $\langle \pi^+ \pi^- | \theta_{\mu}^{\mu} | 0 \rangle $
change the shape of $m_{\pi \pi}$ distribution dramatically.


\section{Decays of $B_c$ into a lepton pair
         combined with two pions}
The observation of the meson $B_c$ via the decay mode $B_{c}^{\pm} \to \jpsi
+ \ell^{\pm} \nu$ has been reported recently by
the Collider Detector at Fermilab (CDF) Collaboration\cite{cdf-bc}.
The $B_c^+$ meson is the lowest-mass bound state containing a charm quark and
a bottom antiquark. It has nonzero flavor and can decay only via weak
interaction. Hence it has a very long lifetime,
$\tau (B_c^+) = 0.46_{- 0.16}^{+ 0.18} ({\rm stat.})
\pm 0.03 ({\rm syst.}) {\rm ps}$.  It will offer a new window
for study the weak decay mechanism of heavy flavors and test various
nonperturbative models for bound states.
The leptonic decay of $B_c$ to one heavy meson has been studied
in various models\cite{chang,bc-o}. In this section we study the
leptonic decay of $B_c^+$ into two pions.  The first part of
this section is devoted to leptonic decay of $B_c^+$ into
two hard pions, the decay into two soft pions is studied in the second part.


\subsection{The leptonic decays of $B_c^+$
            combined with two hard pions}
We study this exclusive decay in the rest frame of $B_c^+$:
\beq
 B_c^+ (P) \to l^+ (p_1) + \nu_l (p_2) + \pi^+ (k_{\pi^+})
 + \pi^-(k_{\pi^-}),
\eeq
where $l= e, \mu$, the momenta are indicated in the brackets.
We study the decay in the region where the
two-pion state has a small invariant mass and has a large
total momentum. Similarly as in Sect. 2
the decay amplitude can be factorized, in which
the nonperturbative effect related to $B_c^+$ meson is represented by a
NRQCD matrix element, and that
related to the two pions is represented by the same distribution amplitude
of two gluons in the isoscalar pion pair $\Phi ^G (x_1,\zeta, m_{\pi \pi})$
which is defined in (13).
The S-matrix element for the decay is
\beq
\smatr = i \frac{G_F}{\sqrt{2}} V_{bc} L_{\mu} \cdot \int d^4 z e^{i q \cdot z}
\langle \pi^+ \pi^- | \bar{b} (z) \gamma^{\mu} ( 1 - \gamma^5) c (z) | B_c^+ \rangle ,
\eeq
where $V_{bc}$ is the Cabibbo-Kobayashi-Maskawa matrix element, $c(z)$ and $\bar b(z)$
is the Dirac field
for $c-$quark and for $b$-quark respectively, $q= p_1 + p_2$ and
\beq
L_{\mu} = \bar{u}(p_2) \gamma_{\mu} (1 - \gamma^5) v (p_1),
\eeq
$\bar{u}(p_2)$ and $v (p_1)$ are the spinors of the leptons. Using the method
in Sec. 2,  keeping leading terms in heavy quark expansion and in velocity expansion, we have
\ba
\label{matr-bc}
\smatr &=& \frac{i G_F}{24 \sqrt{2}} V_{bc}  g_s^2 (2 \pi)^4 ~\delta^{4}
(P - k - q) L^{\mu} \cdot
\langle 0 | \chi^{\dagger}_{b} \psi_c | B_c^+ \rangle
\nonumber \\
&& \times \int_0^1  d x_1 \frac{\Phi^G ( x_1, \zeta, m_{\pi \pi} )}
{x_1 (1 - x_1) } \cdot \left[ \frac{1}{2} d_{T}^{\mu_1 \nu_1} \cdot
H_{\mu \mu_1 \nu_1} (P, k, k_1) \right],
\ea
where $\chi^{\dagger}_{b} (\psi_c)$ is the NRQCD field for $\bar{b} (c)$ quark,
$H_{\mu \mu_1 \nu_1} ( P, k, k_1)$ is the hard part of the decay amplitude and can be
calculated perturbatively. We obtain:
\beq
\label{bc-halfdh}
\frac{1}{2} d_{T}^{\mu_1 \nu_1} \cdot H_{\mu \mu_1 \nu_1} (P, k, k_1)
= \frac{8 M_{Bc} P^{\mu}}{( M_{Bc}^{2} - q^2) m_b m_c},
\eeq
where $ L_{\mu} q^{\mu} = 0$ for $m_l = 0$ is used.
The differential decay width can be written as
\beq
\label{bc-dgam}
d \Gamma = \frac{1}{(2 \pi)^8} \cdot \frac {\pi^2}{32} \cdot
\frac{|\vec{k}|}{M_{\psi}} \cdot \beta ~\beta_{l}^{\prime}
~\overline{\sum} |M|^2 d q^2 d m_{\pi \pi}^{2}
d \cos \theta_{\pi} d \cos \theta_l d \phi ,
\eeq
where $\beta_{l}^{\prime} = 1 - m_l^2 /q^2$ is the
velocity of $l^+$ in the
center mass frame of $l^+ \nu_l$.

\begin{figure}[htb]
\centering
\includegraphics[width=8cm,height=8cm]{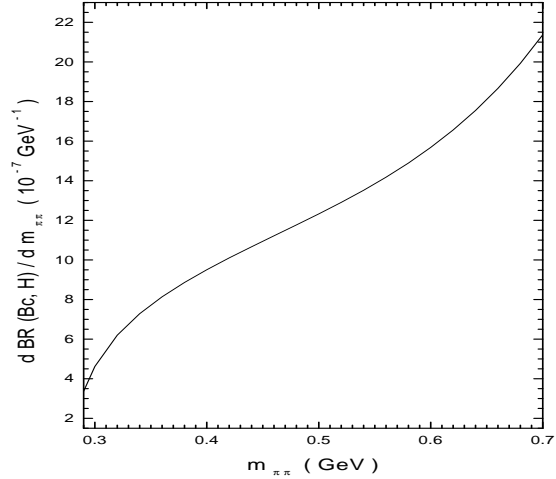}
\vspace*{-15mm}
\caption{The differential decay branching ratio of $B_c^+$,
$d ~{\rm BR} (B_c, H)/dm_{\pi\pi}$ as a function of $m_{\pi\pi}$
in unit of $ 10^{-7} {\rm GeV}^{-1}$ with the cuts.
\label{bchm2pi}}
\end{figure}

\par
To present numerical predictions, the NRQCD matrix element
$\langle 0 | \chi^{\dagger}_{b} \psi_c | B_c^+ \rangle$
should be known. It is related to $B_c$ decay constant $f_{Bc}$
via
\beq
| \langle 0 | \chi^{\dagger}_{b} \psi_c | B_c^+ \rangle |^2
= \frac{1}{2} f_{Bc}^{2} M_{Bc},
\eeq
with $f_{Bc} = 480 {\rm MeV}$ taken from \cite{chang}.
Other parameters take the following values:
$M_{Bc} = 6.4 {\rm GeV}, |V_{bc}| = 4.0 \times 10^{-2},
G_F = 1.166 \times 10^{-5} {\rm GeV}^{-2}, \alpha_s (M_{Bc}) = 0.24$.
We use the cuts: $k^+ \geq 10 ~k^-$, $k^0 + {\bf k} \geq 2.0
{\rm GeV}$ and $2 m_{\pi} \leq m_{\pi \pi} \leq 0.70 {\rm GeV}$.

\begin{figure}[htb]
\centering
\includegraphics[width=8cm,height=8cm]{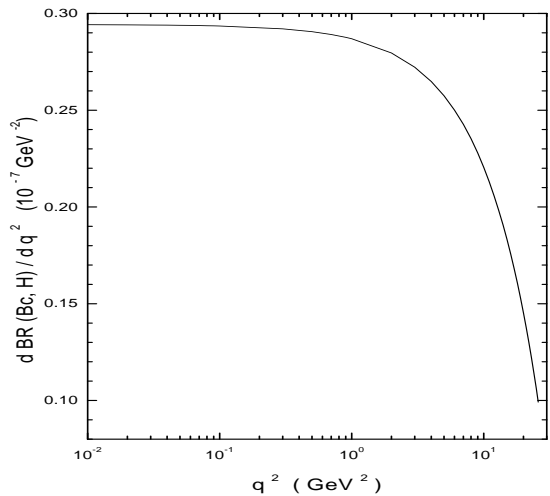}
\vspace*{-15mm}
\caption{The differential decay branching ratio of $B_c^+$,
$d ~{\rm BR} (B_c, H)/d q^2$ as a function of $ q^2$
in unit of $ 10^{-7} {\rm GeV}^{-2}$ with the cuts.
\label{bchqsq}}
\end{figure}

\par
With these parameters and cuts we can predict the differential decay branching
ratio in the considered region.
The $m_{\pi \pi}$ distribution of $B_c^+$ semileptonic decay,
$d ~{\rm BR} (B_c, H)/dm_{\pi\pi}$ in unit of $ 10^{-7} {\rm GeV}^{-1}$
with the cuts
is shown in Fig. \ref{bchm2pi}, the $q^2$ distribution
is presented in Fig. \ref{bchqsq}.
Since the absolute squared matrix element of $B_c^+$ decay
in this region is almost independent of $q^2$, the shape of the $q^2$ distribution
is determined mainly by the phase space factors.
The decay branching ratio is $5.1 \times 10^{-7}$; belonging to it is the s-wave contribution
is $4.7 \times 10^{-7}$. The estimated branching ratio shows that
the decay mode in this region will be not observable even at the Large Hadron
Collier (LHC).


\subsection{The leptonic decay of $B_c^+$
            combined with two soft pions}
In this subsection, we study the leptonic decay of $B_c^+$
combined with two soft pions. We use the same notation for momenta
as before. With the method in \cite{jpma} it is straightforward
to obtain the S-matrix for the decay:
\ba
\label{matr-bcs}
\smatr_s &=& \frac{i G_F}{3 \sqrt{2}} V_{bc} (2 \pi)^4 ~\delta^{4}
(P - k - q)  \langle 0 | \chi^{\dagger}_{b} \psi_c | B_c^+ \rangle
\nonumber \\
&& \times \frac{L_{\mu} \cdot P^{\mu}}{m_b m_c (k^0)^2} T_{\pi \pi} (k),
\ea
where $T_{\pi \pi} (k)$ is defined in  (\ref{smatrix-s}), with our model
for $h(\tau, k)$ given in Sect.3, it can be expressed as
\beq
T_{\pi \pi} (k) = \frac{5 \pi}{k^0} \langle \pi^+ \pi^- | \alpha_s {\bf E}^a (0) \cdot
{\bf E}^a (0) | 0 \rangle.
\eeq

\begin{figure}[htb]
\centering
\includegraphics[width=8cm,height=8cm]{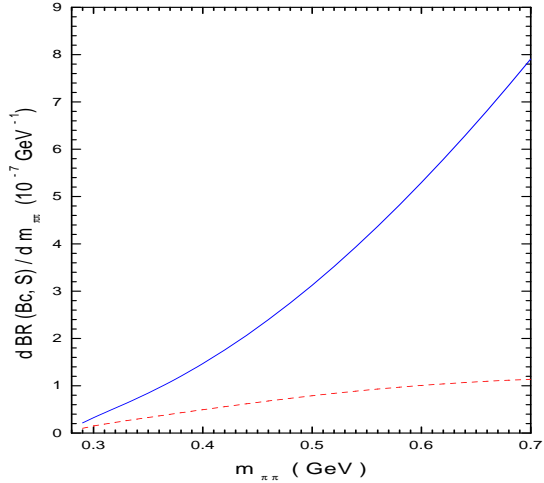}
\vspace*{-15mm}
\caption{The $m_{\pi \pi}$ distribution of $B_c^+$ semileptonic decay to pion pair,
referring as $d ~{\rm BR} (\jpsi, S)/dm_{\pi\pi}$
in unit of $ 10^{-7} {\rm GeV}^{-1}$ with the cuts.
The dashed line denotes the distribution by using the leading order result of
chiral perturbation theory for $\langle \pi^+ \pi^- | \theta_{\mu}^{\mu} | 0 \rangle $,
while the solid line denotes
the distribution by adding one-loop correction
to the matrix element.
\label{bcsm2pi}}
\end{figure}

\noindent
With this S-matrix element, it is straightforward
to obtain the $m_{\pi \pi}$ distribution
of $B_c^+$ decay to two soft pions, which is shown in Fig. \ref{bcsm2pi}.
The cuts are used: $0 \leq |{\bf k}| \leq \frac{1}{10} M_{Bc}$ and
$2 m_{\pi} \leq m_{\pi \pi} \leq 0.70 {\rm GeV}$.
In Fig. \ref{bcsm2pi}, the solid line represents the $m_{\pi \pi}$ distribution
by using next-to-leading order chiral perturbative theory to determine
the matrix element $\langle \pi^+ \pi^- | \theta_{\mu}^{\mu} | 0 \rangle $, while the
dashed line denotes the distribution by using leading order chiral perturbative theory
for this matrix element. Integrating over $2 m_{\pi} \leq m_{\pi \pi} \leq 0.70 {\rm GeV}$,
the decay branching ratios for $B_c^+$ in the kinematic region
are $3.6 \times 10^{-7}$ and $1.5 \times 10^{-7}$, by using the result
at leading- and next-to-leading order of chiral perturbation theory respectively.
The numerical results show that the  decay mode is not observable even at LHC.
But we can learn from  Figs. \ref{psism2pi}, \ref{bcsm2pi}, and  \ref{bbsm2pi}
that when the phase spaces become larger, the next-to-leading order chiral
corrections to the matrix element $\langle \pi^+ \pi^- | \theta_{\mu}^{\mu} | 0 \rangle $
change the shape of $m_{\pi \pi}$ distribution dramatically.


\section{Summary}

In  this paper we have studied the exclusive decay of
$\jpsi, \Upsilon$, and $B_c$ into a lepton pair combined with
two pions, where the two pions can be soft or hard with a small invariant mass.
In both cases the decay amplitude can be factorized,
in which  the nonperturbative effect related to the heavy meson is
represented by a NRQCD matrix element, and that related
to the two pions is represented by a distribution amplitude of two
gluons in the isoscalar pion pair in the case with hard pions, and
by a correlator of chromoelectric fields in the case with soft pions.
With suitable models for gluon conversion into soft or hard pions
we are able to predict branching ratios and different distributions.

\par
Our numerical results show that the leptonic decay of $\jpsi$
combined with two hard pions or with two soft pions can be observed
at BES II and at the
proposed BES III and CLEO-C, while the other decays have a too small
branching ratio to be observed. If the decays of $\jpsi$ are observed
in experiment, it will provide information how gluons, which are fundamental
degrees of freedom in QCD, are converted into observed pions.

\vskip 10mm
\begin{center}
{\bf\large Acknowledgments}
\end{center}

The work of J. P. Ma is supported  by National Nature
Science Foundation of P. R. China and by the
Hundred Young Scientist Program of Academia Sinica of P. R. China,
the work of J. S. Xu is supported by the Postdoctoral Foundation of P. R. China and  by
the K. C. Wong Education Foundation, Hong Kong.


\end{document}